	\documentclass[final,3p,twocolumn]{elsarticle}

	\usepackage[english]{babel}
	\usepackage{amssymb}
	\usepackage{amsmath}
	\usepackage{slashed}
	\usepackage{graphicx}
	\usepackage{natbib}
	\usepackage{setspace}
	\usepackage{subfigure}
	\usepackage{appendix}
	\usepackage[pagewise,columnwise]{lineno}

\graphicspath{{Figures/}}

\begin{document}

\begin{frontmatter}

\title{A T-shaped double quantum dot system as a Fano interferometer: interplay of coherence and correlation upon spin currents}

 \author[IFGW]{I. L. Fernandes}
 \author[IFGW]{G. G. Cabrera}
 \ead{cabrera@ifi.unicamp.br}
 \address[IFGW]{Instituto de F\'isica ``Gleb Wataghin", Universidade Estadual de Campinas, UNICAMP, 13083-859, Campinas, SP, Brazil}

\begin{abstract}
Based on Keldysh non-equilibrium Green function method, we have investigated spin current production in a hybrid T-shaped device, consisting of a central quantum dot connected to the leads and a side dot which only couples to the central dot. The topology of this structure allows for quantum interference of the different paths that go across the device, yielding Fano resonances in the spin dependent transport properties. Correlation effects are taken into account at the central dot and handled within a mean field approximation.  Its interplay with the Fano effect is analyzed in the strong coupling regime. Non-vanishing spin currents are only obtained when the leads are ferromagnetic, the current being strongly dependent on the relative orientation of the lead polarizations.  We calculate the conductance (spin and charge) by numerically differentiating the current, and a rich structure is obtained as a manifestation of quantum coherence and correlation effects. Increase of the Coulomb interaction produces localization of states at the side dot, largely suppressing Fano resonances. The interaction is also responsible for the negative values of the spin conductance in some regions of the voltage near resonances, effect which is the spin analog of the Esaki tunnel diode. We also analyze control of the currents via gate voltages applied to the dots, possibility which is interesting for practical operations.

\end{abstract}



\end{frontmatter}

\section{Introduction}

During the last decade, abundant research has been conducted, both theoretically and experimentally, on spin dependent transport properties in hybrid nanosystems with quantum dots (QDs) in their structure. A sample of representative papers is given in \cite{PhysRevLett.102.236806,PhysRevB.75.165303,Science.306.86,PhysRevLett.104.036804}, and references therein. The growing interest is twofold, basic research and potential applications in new spintronic devices \cite{Spintronics-Awschalom, Nature.3.153.2007}. QDs with a few number of electrons simulate artificial atoms, and as such, they display charge energy effects when extra electrons are added to the dot, promoting some electrons to higher level states. They are excellent prototypes to study electron-electron (e-e) correlations in confined systems, and one can probe fundamental many-body effects when QDs are coupled to charge reservoirs (leads). By applying gate voltages to the dots, one can tune the dot barrier height. For very high barriers (weak coupling), the transport is dominated by electron-electron interactions in the so called Coulomb-blockade regime, where the transport is suppressed unless energy is provided to overcome the Coulomb repulsion when adding an extra electron to the dot.
At intermediate coupling between the QD and the leads, one decreases the confinement of the barriers, and tunneling effects and spin interactions dominate over the Coulomb interactions. The transmission through the dot broadens, leading to Kondo resonance peaks. The effect has been observed when the dot develops a net spin due to odd-electron occupancy \cite{Nature.391.156,Cronenwett24071998}. The signature of the effect is the unitary limit of the conductance at low temperatures and zero bias, \emph{i.e.} $G=G_0$, with $G_0$ being the quantum of conductance. This Kondo resonance would be strongly modified when quantum interference is allowed in the nano-structure. This is the case in systems, whose topology allows the interference of a ballistic channel with the resonant channel from the QD. Destructive interference suppresses the transmission, transforming the Kondo peak into a Kondo valley, which ideally reaches the anti-unitary limit $G=0$ at the anti-resonance position. This situation is understood as a purely quantum phenomenon that results from the interplay of Kondo correlations and quantum interference effects \cite{antikondo,fano-kondo3}, and sometimes is referred in the literature as the Fano-Kondo resonance \cite{fano-kondo1, fano-kondo2}.
If, in the above device, one further decreases the confinement of the dot barrier, the total transmission shows asymmetric peaks and dips, which are universally observed whenever a resonant and non-resonant channels are coherently coupled. Quantum interference of electron waves gives now rise to the conventional Fano effect \cite{PhysRev.124.1866}, where the conductance displays a typical asymmetric line shape near resonances, as function of the voltage. This shape of the conductance can be understood in terms of single-electron physics, and the basis of the phenomenon is similar to the quantum interference in the double-slit experiment \cite{Merzbacher}. In the present paper, we will address ourselves to this particular regime \cite{PhysRevB.62.2188,wu-cao,racec}.


Compared with single-QD systems, multiple QDs may show much more interesting features, allowing for versatile hybrid devices suitable of interesting applications. Particular examples are presented in \cite{RevModPhys.75.1, PhysRevB.69.245327,PhysRevB.72.045332, PhysRevB.82.165304,barnas}, which is far from being a comprehensive list of contributions in this vast field.

\begin{figure}[h!]
\centering
\includegraphics[scale=0.48]{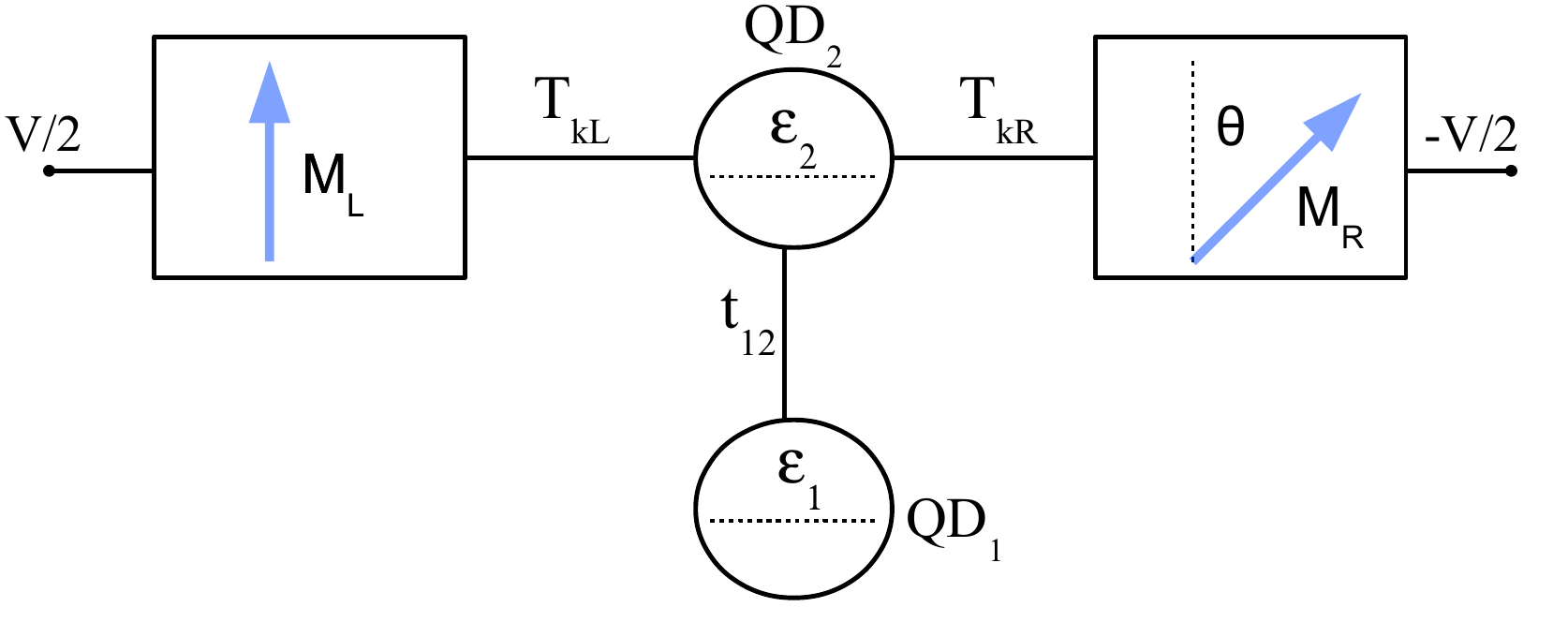}
\caption{Schematic illustration of a T-shaped double quantum dot system with ferromagnetic electrodes. As indicated in the figure, $T_{k\alpha}$ is the coupling between the $\alpha$-electrode and QD$_2$ and $t_{12}$ is the interdot coupling. Magnetization of the leads are indicated by the arrows.}
\label{fig:Diagrama_QD_T}
\end{figure}

In this work, we consider the configuration shown in Fig. \ref{fig:Diagrama_QD_T}, with two quantum dots. A central dot (QD$_2$) is coupled to the leads (source and drain), and to a side dot (QD$_1$), which otherwise will be isolated. This T-shape device is a prototype to study the Fano effect, since the geometry provides an additional path via the side dot, which interferes with the path that goes directly to the drain through the central dot. Low temperatures are required, since the effect critically depends on the preservation of quantum coherence. Transport properties of T-shape structures have been previously studied in a number of papers \cite{Eur.Phys.J.B.79.455,t-shape1,t-shape2,t-shape3} in different contexts. Here, the emphasis is centered on spin currents.

To make the model more realistic, we have also included electronic correlations at the central dot. Electron-electron scattering is a well known source of decoherence, and as such, electronic correlations will compete with the Fano effect. In this contribution, we want to assess the interplay of both phenomena on the transport properties of a nanostructure \cite{PhysRevLett.93.106803}. In addition, we assume the electrodes to be ferromagnetic, and coupled to the central T-shape double QD system via tunneling barriers. Due to spin dependent scattering, tunneling magnetoresistance (TMR) effects are obtained, and one can separate the contributions of the different spins in the description of transport properties.
This is very convenient in order to conceive new and innovative spintronic devices based on spin current phenomena.

As a summary, the system we study theoretically, embodies a number of interesting features, including TMR, Fano resonances, and Coulomb charge effects on spin transport properties. Due to its unique behavior, the device can be used as an interferometer and/or as a spin diode, under the control of gate voltages and magnetic fields. From the experimental side, the generation of spin currents using different methods is nowadays standard, but their detection has remained limited to indirect measurements, as for example, measuring the reorientation of a film magnetization caused by the spin-transfer torque effect. Recently, Zi Qui and collaborators have proposed a method based on X-ray pulses, that directly probes the flow of spin currents as they propagate through the different layers of the sample~\cite{spin.current}, thus avoiding ambiguities usually present when using indirect techniques.

We now comment on the organization of our paper.
In Section \ref{Sec-Model} we describe the model and introduce the theoretical framework, including the Keldysh's non-equilibrium Green function method \cite{Keldysh,GreenFunction-Haug}. In Section \ref{Sec-Results}, numeric calculations of different examples are presented and discussed. Finally, Section \ref{Sec-conclusions} summarizes the results and outcomes of our work.

\section{Model and Method} \label{Sec-Model}
We briefly describe the features of the setup depicted in Fig. \ref{fig:Diagrama_QD_T}. The magnetic polarization of the left lead $M_L$ is kept fixed, while the one on the right $M_R$ is rotated by $\theta$ in relation to the polarization of the first. In our configuration, the central dot, QD$_2$, is connected to the ferromagnetic leads via tunneling coupling $T_{k\alpha}$ ($\alpha$ = L,R, meaning left and right), whereas the side dot, QD$_1$, is coupled to the central dot via a direct hopping term, $t_{12}$. The central dot is assumed to be interacting, with on-site energy $U$. The side dot is noninteracting, and has to be thought as a more extended electron droplet, with a larger capacitance. Its main role is supplying an additional channel for conduction. For simplicity, a single energy level $\varepsilon_i$ ($i=1,2$) has been assigned to each QD$_i$, and the Stoner model is used to describe ferromagnetism at the leads. The applied voltage is $V/2$ and $-V/2$ for the left and right electrodes, respectively. If the correlation $U$ in QD$_2$ is treated within the Hartree-Fock approximation, our system is described by the following model Hamiltonian:

\begin{equation}
H = H_L + H_R + H_{DQD} + H_T,
\end{equation}
\noindent where
\begin{equation}
H_L = \sum_{k\sigma} \left[  \varepsilon_{kL\sigma} - \sigma M_L -\mu_L\right] a^\dag_{kL\sigma}a_{kL\sigma},\nonumber
\end{equation}
\begin{eqnarray}
H_R = \sum_{k\sigma} \{\left[ \varepsilon_{kR\sigma} - \sigma M_R \cos \theta - \mu_R \right] a^\dag_{kR\sigma}a_{kR\sigma} \nonumber \\  - M_R \sin \theta a^\dagger_{k R \sigma}a_{k R\overline{\sigma}}\},\nonumber
\end{eqnarray}
\begin{equation}
H_{DQD}  = \sum_{\sigma} [E_{2\sigma}d^\dag_{2\sigma}d_{2\sigma} + \varepsilon_{1}d^\dag_{1\sigma}d_{1\sigma}]\nonumber
\end{equation}
\begin{equation}
H_T = \sum_\sigma t_{12}d^\dag_{1\sigma}d_{2\sigma}
 +  \sum_{k\sigma\alpha} T_{k\alpha}a^\dag_{k\sigma\alpha}d_{2\sigma} +H.c. \nonumber
\end{equation}
$H_L$ and $H_R$ are the free Hamiltonians of the left and right ferromagnetic electrodes, $a^\dag_{k\alpha\sigma}$ is the creation operator of electrons with spin $\sigma$ and wave vector $\mathbf{k}$ at the lead $\alpha$ ($\alpha = L,R$), and $\varepsilon_{k\beta\sigma}$ is the corresponding single-electron band energy. Within the Stoner model, spin bands are rigidly shifted by the internal magnetization. The magnetization direction of the left electrode is chosen to quantize the spin, and the magnetization of the right one is rotated by $\theta$ with respect to the left one. Notice that $H_R$ is not diagonal in spin due to the above rotation. $H_{DQD}$ is the Hamiltonian for the double quantum-dot system within the Hartree-Fock approximation for QD$_2$. It is written as an effective free Hamiltonian, where $E_{2\sigma} = \varepsilon_2 + U \left<\hat{n}_{2\overline{\sigma}}\right>$ is the energy level in QD$_2$, renormalized by the Coulomb interaction $U$ and the mean occupation of the opposite spin. In the above formulae, $d^\dag_{n\sigma}$, with $n=1,2$, is the electron creation operator at the QD$_n$. We assume that the inter-dot Coulomb repulsion is small and can be neglected. $H_T$ is the transfer Hamiltonian between dots and between the leads and QD$_2$. It describes the tunneling between dots with amplitude $t_{12}$ and tunneling between the electrode $\alpha$ and QD$_2$ with amplitude $T_{k\alpha}$.
To fix ideas, we calculate the current $I_{L\sigma}$ ($\sigma=\uparrow,\downarrow$) at the left electrode. The electric and spin currents, $I_e$ and $I_s$ respectively, are obtained as:
\begin{eqnarray}
I_e & = & I_{L\uparrow} + I_{L\downarrow} \label{eq:Ie}, \\
I_s & = & I_{L\uparrow} - I_{L\downarrow} \label{eq:Is}.
\end{eqnarray}

Calculation of the current is done using the equation $I_{L\sigma} = - e <\dot{N}_{L\sigma}>$, where $N_{L\sigma} = \sum_ka^\dagger_{k L \sigma}a_{k L \sigma}$ is the number operator. Heisenberg equation $\dot{N}_{L\sigma} = i/\hbar \left[H,N_{L\sigma}\right]$ leads to:
\begin{equation}
I_{L\sigma} = - \frac{2e}{\hbar}  \Re \left[ \sum_{k} \int \frac{d\epsilon}{2\pi} T_{kL\sigma}^* G_{kL,2}^{\sigma\sigma,<}(\epsilon) \right], \label{current}
\end{equation}

\noindent where $G_{kL,2}^{\sigma\sigma,<}(\epsilon)$ is the Fourier transform of the lesser Green function $G_{kL,2}^{\sigma\sigma,<}(t,t') = i \left< a_{k\sigma L}\left(t\right)d^\dagger_{2\sigma}\left(t'\right)\right>$ in the Keldysh formalism.

Applying the Langreth theorem \cite{GreenFunction-Haug} and taking the Fourier transform, the lesser Green function is written as:
\begin{equation}
\mathbf{G}^{<}_{kL,2} (\epsilon) =  \mathbf{g}^{r}_{kL} (\epsilon) \mathbf{T}_{kL} \mathbf{G}^{<}_{2} (\epsilon) + \mathbf{g}^{<}_{kL} (\epsilon) \mathbf{T}_{kL} \mathbf{G}^{a}_{2} (\epsilon),
\label{eq:green_lesser_kL2_matriz}
\end{equation}

\noindent where the labels $r$, $a$ and $<$ mean retarded, advanced and lesser, respectively. $\mathbf{G}^{a(<)}_{2}$ is the Green function of QD$_2$, $\mathbf{g}^{r(<)}_{kL}$ is the Green function of the left electrode and $\mathbf{T}_{kL}$ describes the coupling between the dot QD$_2$ and the left electrode. The Green function of the central quantum dot QD$_2$ includes the Coulomb correlation at the dot and the coupling with the other dot QD$_1$. In spin space, the matrix of the Green function can be written as:
\begin{eqnarray}
\mathbf{G}^{s (<)}_{}  = \left( \begin{array}{cc}
G^{\uparrow\uparrow, s (<)}_{} & G^{\uparrow\downarrow ,s (<)}_{}  \\
G^{\downarrow\uparrow s (<)}_{}  & G^{\downarrow\downarrow ,s (<)}_{}
\end{array} \right),
\end{eqnarray}

\noindent where $s=a,r$. The Fourier transform of the left electrode Green functions are given by:
\begin{equation}
g_{kL}^{\sigma ,<} = i 2\pi
f_L(\epsilon_{k\sigma L}) \delta(\epsilon-\epsilon_{k\sigma L} ) \label{eq:G_lesser_sigma_FM_isolado},
\end{equation}
\begin{equation}
g_{kL}^{\sigma ,r(a)} =
\frac{1}{\epsilon-\epsilon_{k\sigma L} \pm i\eta},
\label{eq:G_retardada_sigma_FM_isolado}
\end{equation}

\noindent where $f_\alpha$ is the Fermi distribution of the electrode $\alpha$.
The lesser Green function of the quantum dot $QD_2$ can be calculated by the Keldysh equation $\mathbf{G}^<_2 = \mathbf{G}^r_2\ \mathbf{\Sigma}^<_T\ \mathbf{G}^a_2$. Thus, defining $\mathbf{A} =  \mathbf{\Gamma}_L\left(\epsilon\right) \mathbf{G}^{r}_{2}\left(\epsilon\right)\mathbf{R} \mathbf{\Gamma}_R \left(\epsilon\right)\mathbf{R}^\dagger \mathbf{G}^{a}_{2}\left(\epsilon\right)$, after some algebraic manipulation  we can simplify $I_\sigma$ to:

\begin{equation}
I_\sigma =  \frac{e}{\hbar}  \int \frac{d\epsilon}{2\pi} \left[f_L\left(\epsilon\right) - f_R\left(\epsilon\right) \right] A_{\sigma\sigma}\left(\epsilon\right),
\end{equation}
noting that $\mathbf{A}$ is a 2x2 matrix in spin space,
\begin{equation}
\mathbf{A} = \left( \begin{array}{cc}
A_{\uparrow\uparrow}  & A_{\uparrow\downarrow}  \\
A_{\downarrow\uparrow}  & A_{\downarrow\downarrow}
\end{array} \right)
\end{equation}
with the $\mathbf{R}$ matrix describing the rotation of the magnetization from the left to the right electrode:
\begin{equation}
\mathbf{R} = \left( \begin{array}{cc}
\cos\frac{\theta}{2}  & -\sin\frac{\theta}{2}  \\
\sin\frac{\theta}{2}  & \cos\frac{\theta}{2}
\end{array} \right).
\end{equation}

The coupling $\mathbf\Gamma_{\alpha}(\epsilon)$ is the diagonal matrix whose elements are given by $\Gamma_{\alpha\sigma}(\epsilon) = 2\pi\sum_{k}|T_{k\alpha}|^2\delta(\epsilon-\epsilon_{k\sigma \alpha})$. We assume that the coupling strength is in the wide-band limit, and therefore, $\Gamma_{\alpha\sigma}$ is considered as constant (with a constant band density of states). Under these considerations, the tunnel current and the spin current have the forms of:
\begin{eqnarray}
I_e &=&  \frac{e}{\hbar}  \int \frac{d\epsilon}{2\pi} \left[f_L\left(\epsilon\right) - f_R\left(\epsilon\right) \right] \text{Tr}\ \mathbf{A},\label{e_current} \\
I_s &= & \frac{e}{\hbar}  \int \frac{d\epsilon}{2\pi} \left[f_L\left(\epsilon\right) - f_R\left(\epsilon\right) \right] \text{Tr}\left[\sigma_z \mathbf{A} \right],\label{s_current}
\end{eqnarray}
When both electrodes are identical (same polarization), spin dependent couplings from the left and right are equal (symmetric coupling), and $G^{<}$ can be eliminated in the expression of the current. This way, Eqs. (\ref{e_current}) and (\ref{s_current}) are given in terms of $G^{r}$ and $G^{a}$ only \cite{proportional}. When the leads are different and the dots have more than one level, $G^{<}$ has to be obtained using other methods \cite{nonproportional}.\\

Using the equation of motion approach \cite{Keldysh, GreenFunction-Haug}, and adopting the mean-field approximation to treat the Coulomb interaction at QD$_2$, the Green functions can be put into closed form.  The retarded Green function is given by:
\begin{equation}
\mathbf{G}_{2}^r \left(\epsilon\right) = \mathbf{g}_{2}^r \left(\epsilon\right) \left[1-\mathbf \Sigma_t^r\ \mathbf{g}_{2}^r \left(\epsilon\right)\right]^{-1}
\end{equation}

\noindent where $\mathbf{g}_{2}^r$ is the Green function for the isolated QD$_2$ and $\mathbf\Sigma_t^r$ is the retarded self-energy given by $\mathbf\Sigma_t^r = -i \left(\mathbf\Gamma_{L} + \mathbf{R}\ \mathbf\Gamma_{R}\ \mathbf{R}^\dagger + \mathbf\Gamma_{12} \right)/2$, with $\mathbf\Gamma_{12}$ being the coupling between dots.

Since the Green function depends on the mean occupation via the energy $E_{2\sigma}$, average values are calculated self-consistently, as shown below:
\begin{equation}
\left< n_{2\sigma}\right> = -i \int \frac{d\epsilon}{2\pi}   G_{2}^{\sigma\sigma,<}(\epsilon).
\end{equation}
Once the occupation number is obtained as a function of the bias voltage, we determine the electric and spin currents (Eq.\ref{e_current} and Eq.\ref{s_current}). In particular, the spin dependent Local Density of States (LDOS) at the quantum dots is given by the imaginary part of the corresponding retarded Green function:
\begin{equation}
D^{\sigma}_\lambda(\epsilon)=-\frac{1}{\pi}~\Im~G^{\sigma\sigma,r}_\lambda (\epsilon),
\label{ldos}
\end{equation}
with the index $\lambda=1,2$ denoting the dot. Thus, the total LDOS at the quantum dot is given by:
\begin{equation}
D_\lambda(\epsilon;V)=D^{\uparrow}_\lambda (\epsilon;V) + D^{\downarrow}_\lambda (\epsilon;V)~.
\label{total}
\end{equation}
Since our system is stationary but out of equilibrium, quantities of interest are voltage dependent, including the LDOS of (\ref{ldos}) and (\ref{total}). However, for fixed $U$, our calculation shows that the LDOS are weakly dependent on voltage, with tiny differences related to a small polarization of the dots. This fact will be commented later on.

\section{Results and discussion} \label{Sec-Results}

In the following, we present numerical results. To simplify, we assume that electrodes are made of the same material, e.g., $P_R = P_L = P$, where $P_\alpha$ is a generalized polarization of the $\alpha$-electrode, defined as $P_{\alpha} = \left(\Gamma_{\alpha\uparrow} - \Gamma_{\alpha\downarrow}\right) / \left(\Gamma_{\alpha\uparrow} + \Gamma_{\alpha\downarrow}\right)$. Therefore, we get $\Gamma_{\alpha \sigma} = \Gamma_{\alpha}(1 + \sigma P_{\alpha})$, with $\Gamma_{\alpha} = \left(\Gamma_{\alpha\uparrow} + \Gamma_{\alpha\downarrow}\right)/2$. All the energies are given in units of $\Gamma_0 = \Gamma_{L} = \Gamma_{R}$, which defines our energy scale for the Fano regime. Most of the examples are calculated for fixed values of $\epsilon_1 = \epsilon_2 = 3\ \Gamma_0$ and $k_bT = 0.03\ \Gamma_0$, unless otherwise stated. Finally, the external voltage $V$ applied across the system is given by $eV/2=\mu_L\ =-\mu_R\ $, where $\mu_\alpha$ is the chemical potential of the $\alpha$-electrode ( $\alpha=L,R$).

\begin{figure}[h!]
\centering
\includegraphics[scale=0.45]{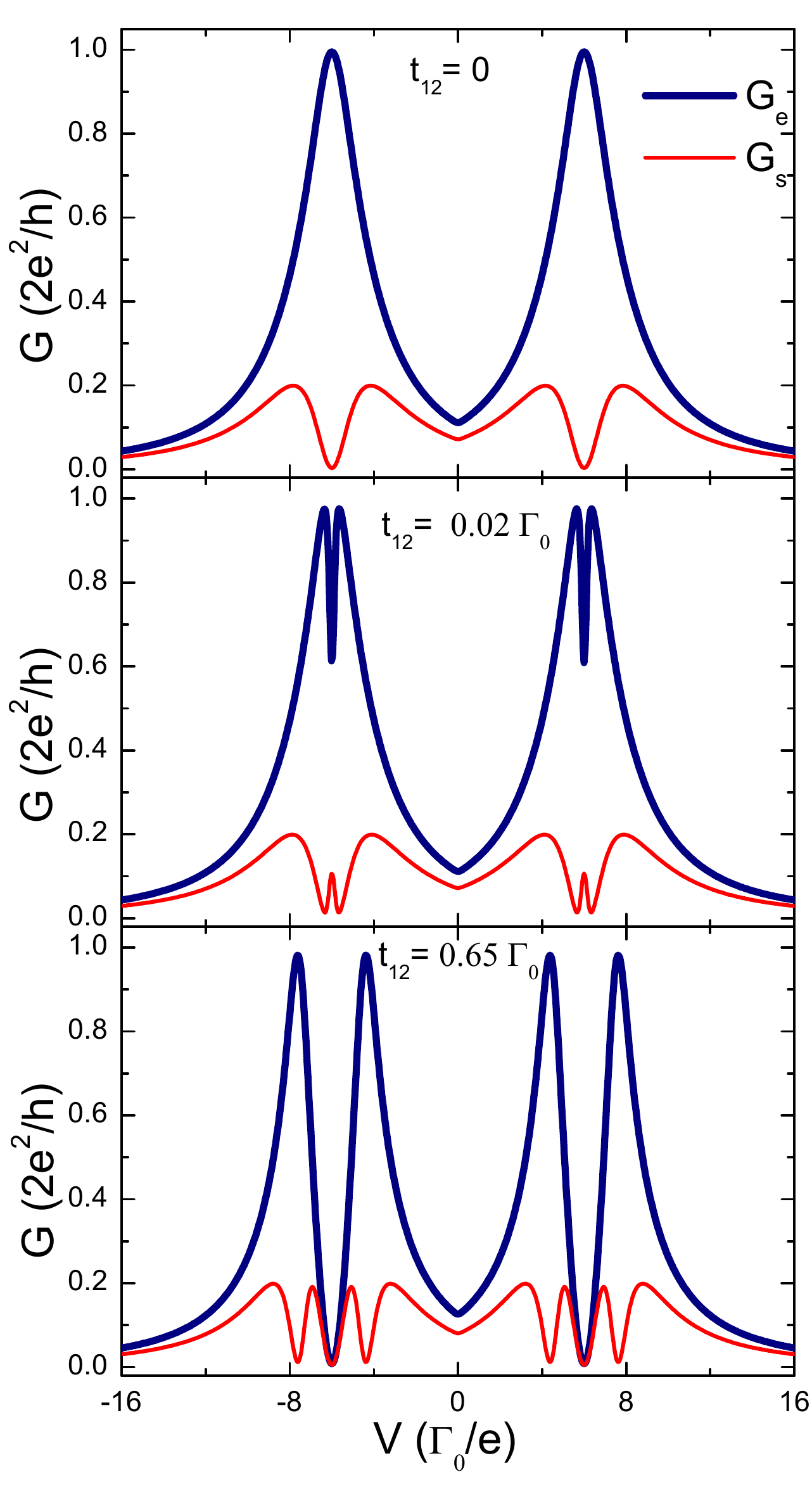}
\caption{The bias dependance of G$_e$ (wide blue line) and G$_s$ (fine red), for different values of the interdot coupling, in the parallel configuration of the magnetization, with $P=0.4$. Other parameters are $k_bT = 0.03\ \Gamma_0$, $\epsilon_1 = \epsilon_2 = 3$ $\Gamma_0$ and $U = 0$.}
\label{fig:gra_G_e-s_V_theta-0_t12-0-065-2}
\end{figure}

\subsection{Non-interacting case}

The bias dependence of the differential conductance, defined by $G_r = dI_r/dV$, with $r=e,s$, is studied. It was calculated by numerically differentiating the current (Eq. \ref{eq:Ie} and \ref{eq:Is}) with respect to the applied voltage. Fig. \ref{fig:gra_G_e-s_V_theta-0_t12-0-065-2} shows the electric and spin differential conductance as a function of the bias for different couplings between dots, in the parallel configuration ($\theta=0$), with $P = 0.4$. Energy levels for the dots are fixed with values $\epsilon_1 = \epsilon_2 = 3$ $\Gamma_0$. When the side dot is isolated ($t_{12} = 0$), the current goes directly through the central dot, and the electrical conductance displays a Breit-Wigner line shape with resonances at $V = \pm 6.0$ $\Gamma_0/e$, as shown in the figure. The width of the resonance is due to coupling of the central dot with the leads. Note that for this particular configuration, with $V = 6.0$ $\Gamma_0/e$, the dot levels are aligned with the chemical potential of the left electrode, and the conductance is at a maximum. A symmetric situation is obtained when $V = -6.0$ $\Gamma_0/e$, with the levels aligned with the right chemical potential. For non vanishing $t_{12}$, G$_e$ develops a behavior characteristic of Fano resonances. Interference effects give rises to Fano peaks and dips, with the Breit-Wigner peaks splitting into two Fano peaks and a dip. The splitting of the peaks is roughly given by $2 t_{12}$, with a factor $1/2$ to scale the voltage with the energy (due to the symmetric choice to measure the voltage). Thus, when the coupling between dots increases, the splitting between Fano peaks also increases, thus broadening the resonance. The Fano dips instead are pinned at $V = \pm 6$ $\Gamma_0/e$, where the two channels through both QD's are open and interfere, but the shape of the resonance is strongly affected by the coupling between dots. For large enough $t_{12}$ the anti-resonance is complete ($G_e=0$).


 The overall behavior of the spin conductance shows more structure than G$_e$. Assuming that $P > 0$, we have $\Gamma_{\alpha\uparrow} > \Gamma_{\alpha\downarrow}$ for both ferromagnetic electrodes. This imbalance of the tunneling couplings induces a spin current flowing through the T-shaped junction. When the side dot is isolated, the spin conductance displays a dip exactly at the position of the peak for $G_e$. This pattern is maintained with the multiple splitting of the Breit-Wigner line shape, when interference is present. Several peaks and dips are generated, with a Fano antiresonance always pinned at $V = \pm 6$ $\Gamma_0/e$. Since $G_s$ is always positive, the spin current increases with the voltage, displaying a small plateau anytime that G$_s$=0. No spin current is present for the unpolarized case ($P=0$).


\subsection{Interacting case}

\begin{figure}[h!]
\centering
\includegraphics[scale=0.52]{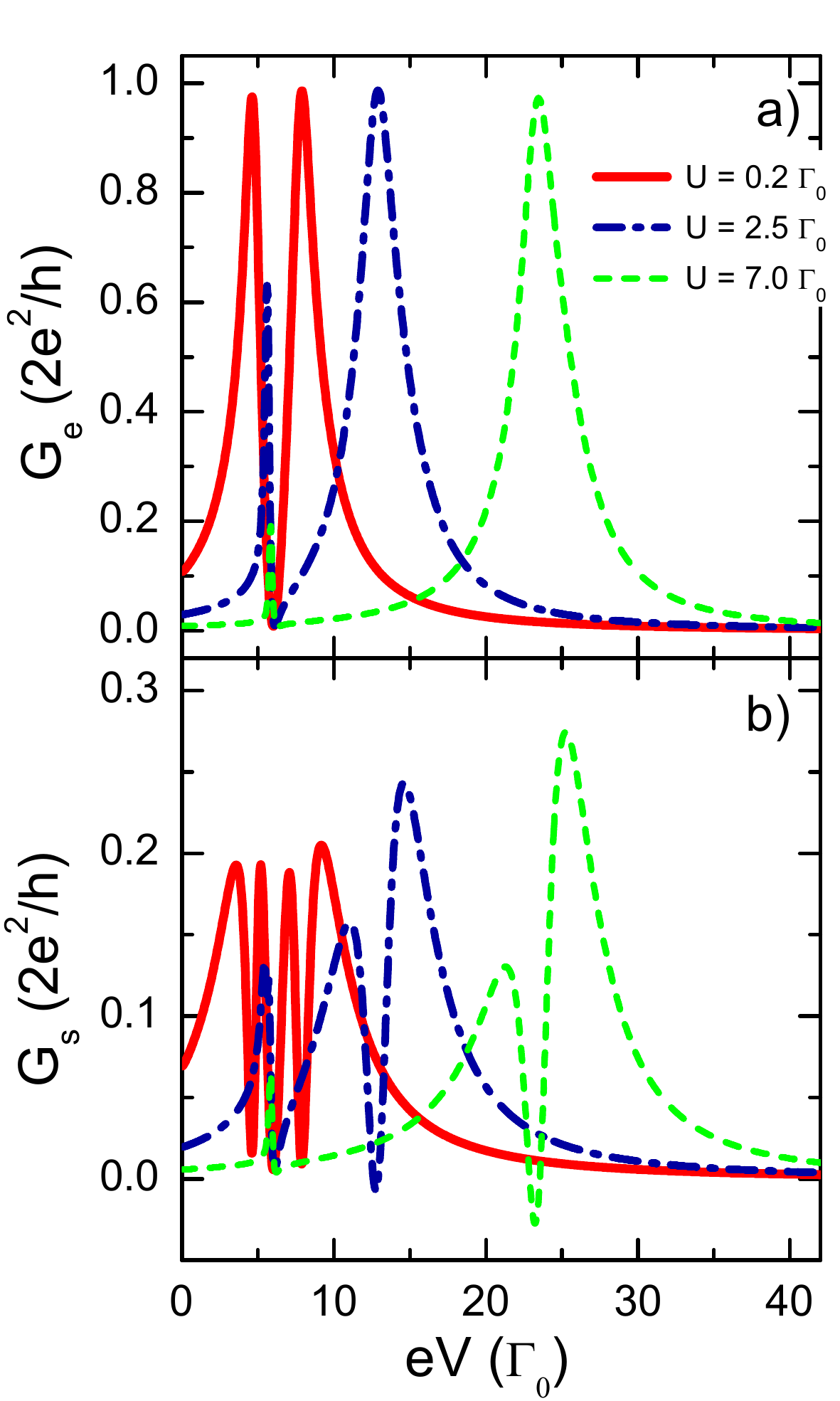}
\caption{The bias dependence of the electric and spin conductances, G$_e$ and G$_s$, as a function of the applied voltage, for different values of Coulomb interaction $U$: a) G$_e$ at $P=0$ and $t_{12} = 0.65$; b) G$_s$ for the parallel configuration of magnetizations, with $P=0.4$ and $t_{12} = 0.65$.
Other parameters are $k_bT = 0.03\ \Gamma_0$, $\epsilon_1 = \epsilon_2 = 3$ $\Gamma_0$.}
\label{fig:graf_G-V_U-varia}
\end{figure}

Next, we present mean-field results with the inclusion of the Coulomb interaction $U$ at the central dot. Fig. \ref{fig:graf_G-V_U-varia}a) shows the electric conductance for the unpolarized case, for different values of $U$ and positive voltages. The interdot coupling is $t_{12} = 0.65$, value included in the previous figure. The low voltage Fano peak gets narrower, shifts and gets pinned at the antiresonance position. Its amplitude is suppressed with increasing $U$. The other peak is shifted to higher voltages, and for large $U$ becomes more symmetrical, resembling a Breit-Wigner resonance, with the resonant position being renormalized by the interaction. Perfect Breit-Wigner line shape would indicate that this channel does not participate anymore in interference phenomena. The spin conductance is displayed in Fig. \ref{fig:graf_G-V_U-varia}b), for $P=0.4$, in the parallel configuration. It presents similar general trends: the low voltage Fano peaks converge to the position of the antiresonance, and are suppressed with the interaction. In turn, the high voltage part of G$_s$ presents an interesting behavior. The peaks shift with the interaction and  G$_s$ becomes negative around the position of the resonances of G$_e$. This interesting effect of negative conductance is the spin analog of the tunnel diode phenomena \cite{esaki}. For operational purposes, this condition can be fulfilled, by tuning the voltage at the Breit-Wigner peak of the electric conductance.

\begin{figure}[h!]
\centering
\includegraphics[scale=0.5]{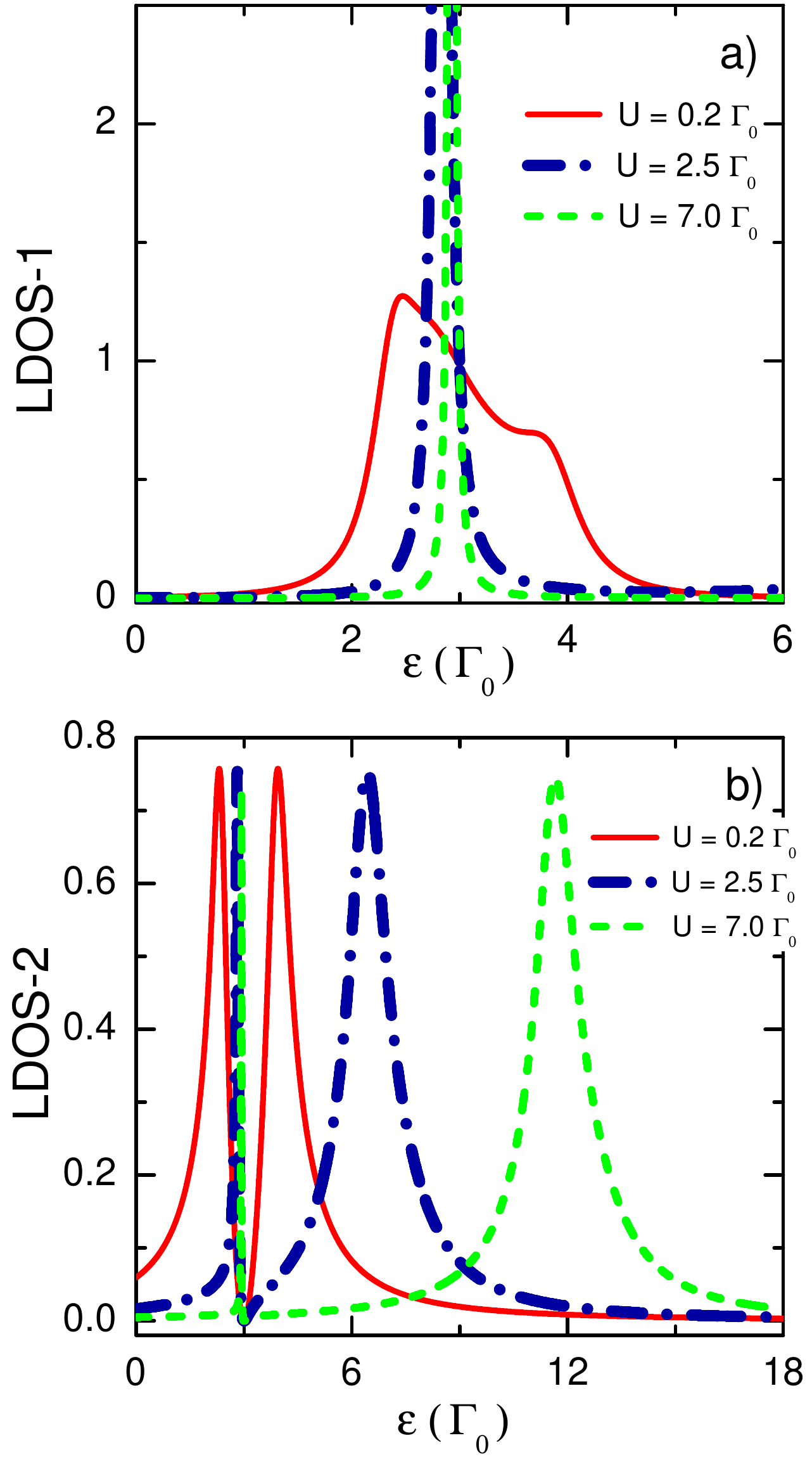}
\caption{LDOS of the two dots, for different values of the Coulomb interaction $U$, for the unpolarized case, $P=0$. a) LDOS for QD$_1$, which is the side dot. The LDOS undergoes a strong localization, with increasing $U$. b) LDOS for QD$_2$, which is the dot that couples directly to the ferromagnetic leads. The interdot coupling is $t_{12} = 0.65\ \Gamma_0$ and the applied voltage $V = 6\ \Gamma_0/e$. Other parameters are $k_bT = 0.03\ \Gamma_0$, $\epsilon_1 = \epsilon_2 = 3$ $\Gamma_0$.}
\label{fig:graf_LDOS_varia_U}
\end{figure}

We have calculated the LDOS at the dots for the various values of $U$, to correlate with the conductance behavior. The discussion that follows is not completely rigourous, since the LDOS are voltage dependent. But, as commented above, this dependence is weak, with small differences associated with the polarization of the dots. For the side dot QD$_1$, the density of states progressively localizes around the dot level, as $U$ is increased, meaning that the dot isolates from the T-shape junction, thus dimming interference effects. This behavior is related with the narrowing of the Fano resonance and the decrease of its amplitude, as shown in Fig. \ref{fig:graf_G-V_U-varia}a). For the central dot QD$_2$, the density of states is split by the interaction into a double peak structure, the low voltage peak being pinned at the unperturbed value $\epsilon_2$, while the second shifts with the correlation $U$. The low voltage peak is the one that participates in the Fano resonance, and becomes more localized as long as $U$ is increased. One sees that LDOS-2 behavior correlates very closely with the electric conductance shown in Fig.~\ref{fig:graf_G-V_U-varia}a), for the same values of parameters.


\begin{figure}[h!]
\centering
\includegraphics[scale=0.40]{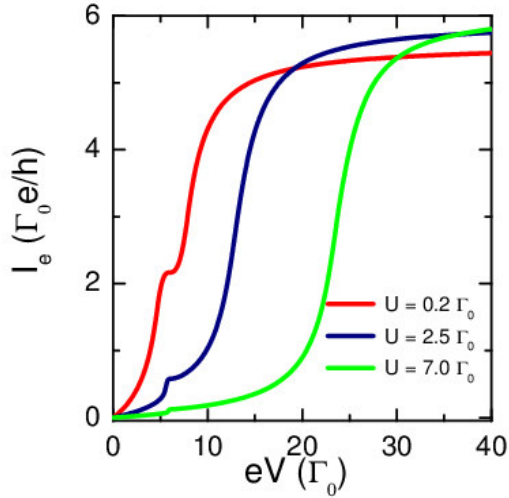}
\caption{The electric current as a function of the applied bias, for different values of the Coulomb interaction. Values of $U$ and other parameters are the same as in the previous figure, in order to correlate their behavior. To scale with the energy, the voltage has to be divided by a factor 2.}
\label{fig:current-with-U}
\end{figure}

\begin{figure}[t!]
\centering
\includegraphics[scale=0.58]{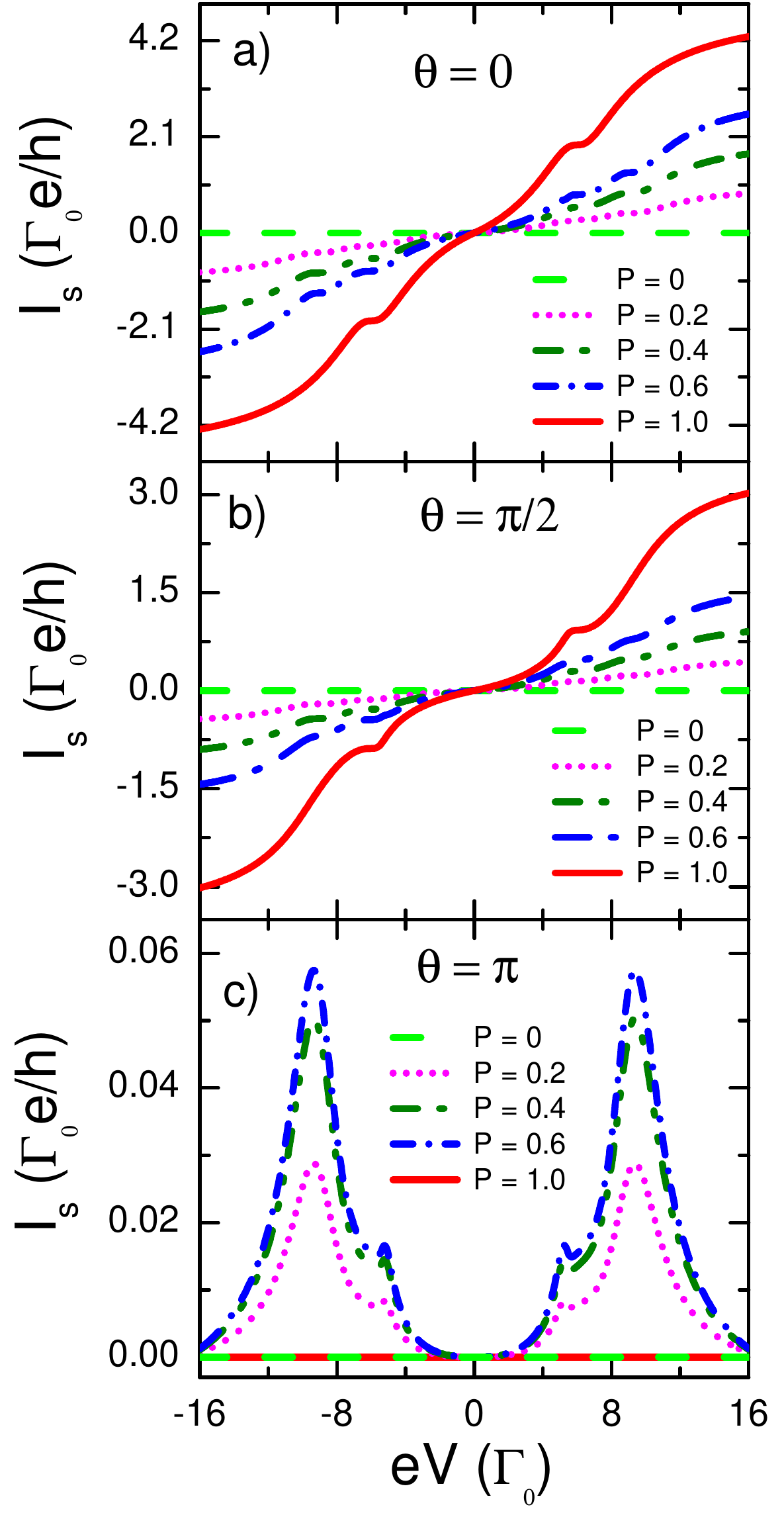}
\caption{The bias dependence of the  spin current for different polarization in a) parallel magnetization configuration, b) for $\theta = \pi/2$ and c) in anti-parallel magnetization configuration. Fixed parameters: $t_{12} = 0.65$ $\Gamma_0$, $k_bT = 0.03$ $\Gamma_0$, $\epsilon_1 = \epsilon_2 = 3$ $\Gamma_0$ and $U = 0.9$ $\Gamma_0$.}
\label{fig:graf_Is-V_theta-pi_U-09_P-varia}
\end{figure}

\begin{figure*}[tpb]
\centering
\includegraphics[width=\textwidth]{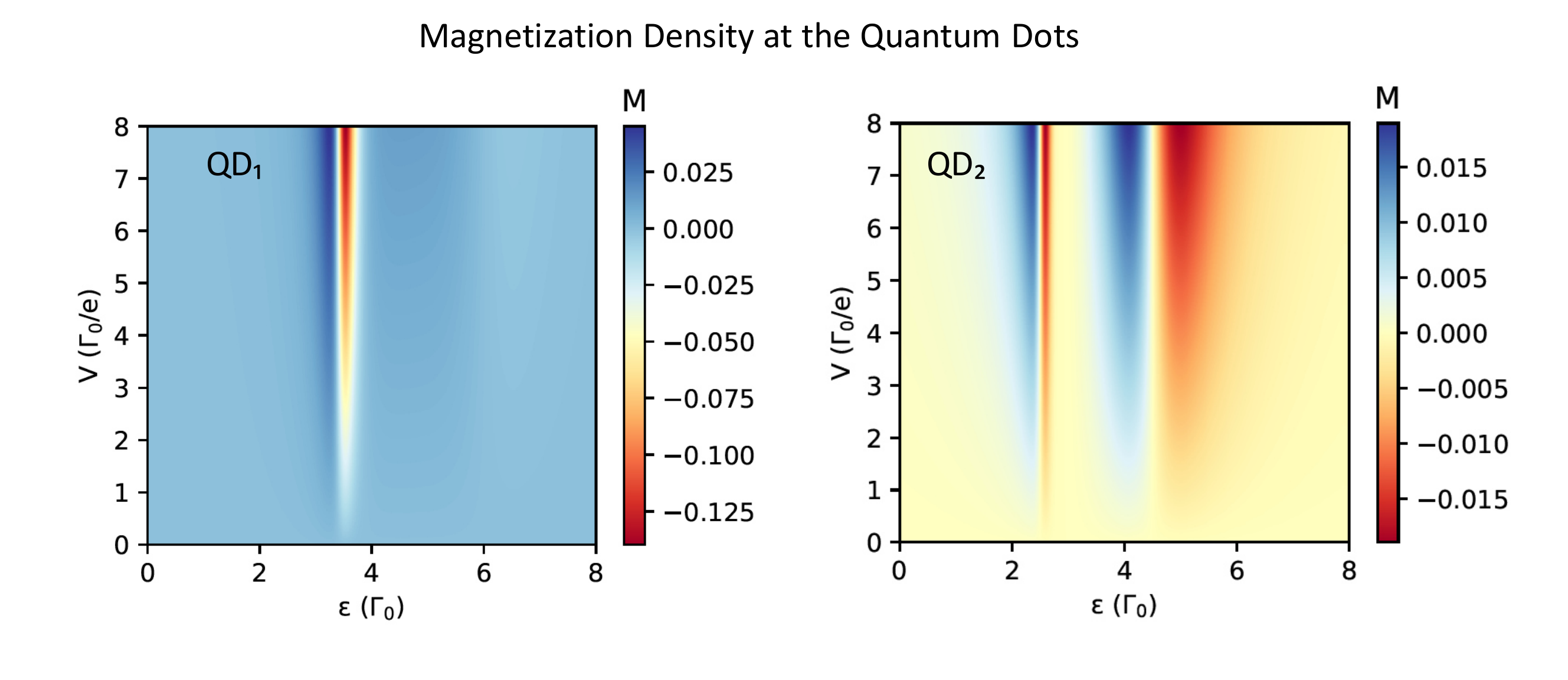}
\caption{Counter plot of the dot magnetization densities, as function of the energy and voltage. The system is in the antiparallel configuration, with $P = 0.6$. Other parameters are $k_BT = 0.03~\Gamma_0$, $\epsilon_1=\epsilon_2 = 3~\Gamma_0$, $U=0.9~\Gamma_0$, and $t_{12} = 0.65~\Gamma_0$~.}
\label{counterplot}
\end{figure*}
In Fig. \ref{fig:current-with-U}, we display the total electric current to illustrate the LDOS features. One observes a significant drop of the current with increasing $U$, up to voltages that open the extra channel corresponding to the second peak of LDOS-2. The Fano resonance appears as a `shoulder' at low voltages, and is less pronounced as the correlation is increased. The steep slope of the curves, after the Fano effect, marks the onset of the Breit-Wigner-like peak shown in Fig. \ref{fig:graf_LDOS_varia_U}b).

In the following, we investigate the bias dependence of the spin current (I$_s$) for several magnetization configurations. The current vanishes identically for the non-magnetic case ($P = 0$). Finite values of $P$ yield a spin current which is susceptible to the relative orientation of the electrode magnetizations, as shown in Fig. \ref{fig:graf_Is-V_theta-pi_U-09_P-varia}. The parallel ($\theta = 0$) and perpendicular ($\theta = \pi/2$) configurations (shown in panels a) and b), respectively), display the same qualitative behavior, with the spin current increasing monotonically with the polarization. We also observe a small plateau pattern due to the Fano anti-resonant tunneling. Much more interesting is the anti-parallel case, shown in Fig. \ref{fig:graf_Is-V_theta-pi_U-09_P-varia}c). For this configuration, the spin populations are interchanged when going from the left to the right electrode, so the system is a `perfect' insulator when $P=1$. We also note that no spin current is obtained for $U=0$, so the non vanishing results shown in the figure is due to the presence of the interaction. The Fano resonance appears as a `shoulder' and a peak, in the $I-V$ characteristics. The current, in all cases, is much smaller than in the previous examples, but its absolute value depends on the constant $\Gamma_0$. Since the system is in the anti-parallel configuration, the small spin current obtained with voltage should be ascribed to magnetism induced by the correlation at the dots, in the presence of polarized leads. In fact, the spin-dependent LDOS at the dots are slightly different for the two spins, producing a voltage dependent magnetic moment at the central part of the T-shape structure. We can define a magnetization density at the dots as:
\begin{equation}
M_\lambda(\epsilon;V)=D^{\uparrow}_\lambda (\epsilon;V) - D^{\downarrow}_\lambda (\epsilon;V)~,
\label{mag}
\end{equation}
with $\lambda=1,2$ as the index for the dots, and the LDOS defined in (\ref{ldos}). Note that $M$ in (\ref{mag}) is voltage dependent and induced by the applied voltage. For one of the examples of Fig. \ref{fig:graf_Is-V_theta-pi_U-09_P-varia}c), we have produced a counter plot of the dot magnetization densities in Fig. \ref{counterplot}, clearly illustrating the regions with non-vanishing values. One observes that $M$ assumes very small values, either positive or negative, in a finite range of the energy. This out-of-equilibrium feature is clearly a many body effect, since it requires, in addition of the correlation, a finite polarization of the leads, with $P\neq 0,1$.

\begin{figure}[t]
\centering
\includegraphics[scale=0.41]{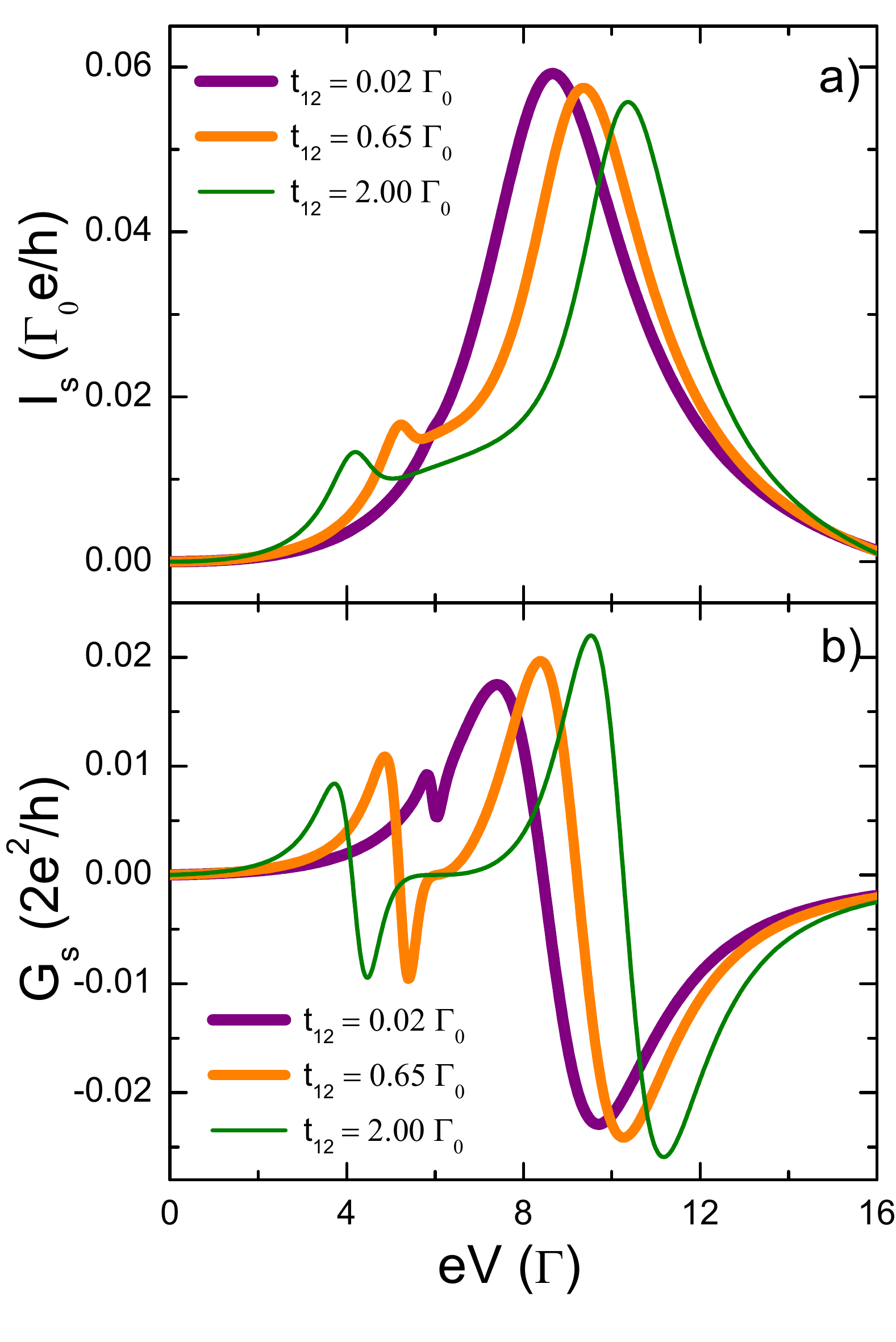}
\caption{The bias dependence of the a) spin current $I_s$ and b) the spin differential conductance G$_s$ for different coupling between the dots $t_{12}$ in anti-parallel configuration. Fixed parameters: $P=0.6$, $k_bT = 0.03$, $\epsilon_1 = \epsilon_2 = 3~\Gamma_0$ and $U = 0.9~\Gamma_0$. Since the current is symmetric, we only plot positive voltages. }
\label{fig:graf_Is-Gs-V_theta-0_U-09-75}
\end{figure}

To resolve the peak structure of $I_s$ for the antiparallel configuration, we have varied the inter-dot coupling $t_{12}$ and calculated the corresponding spin conductance. We observe in Fig. \ref{fig:graf_Is-Gs-V_theta-0_U-09-75}, the presence of two Fano resonances, which for small $t_{12}$, are superposed. With the increase of $t_{12}$, these two resonances separate completely, displaying the typical asymmetric Fano shape in the conductance, as shown in Fig. \ref{fig:graf_Is-Gs-V_theta-0_U-09-75}b). Negative values of the conductance are obtained around resonances, when plotted as functions of the voltage. The $I-V$ characteristics resemble the tunnel diode behavior \cite{esaki}, but this case is even more interesting because there are two windows of voltages with negative conductance. Tuning the voltage around the resonances, one can choose regions of positive or negative spin conductance, with interesting potential applications.

The validity of this effect should be critically discussed, due to limitations of our approach. It is well known that mean field approximations overestimates the formation of magnetic moments \cite{hubbard,anderson}, neglecting magnetic fluctuations. Experimentally, the observation is a challenge due to difficulties in obtaining identical leads. For not too different electrodes, even in the antiparallel configuration, one may get a leak of spin current that will mask the effect. At any rate, this topic deserves further study using approximations more accurate than mean field or by devising an appropriate experimental setup to probe the effect.

\begin{figure*}[tpb]
\centering
\includegraphics[width=\textwidth]{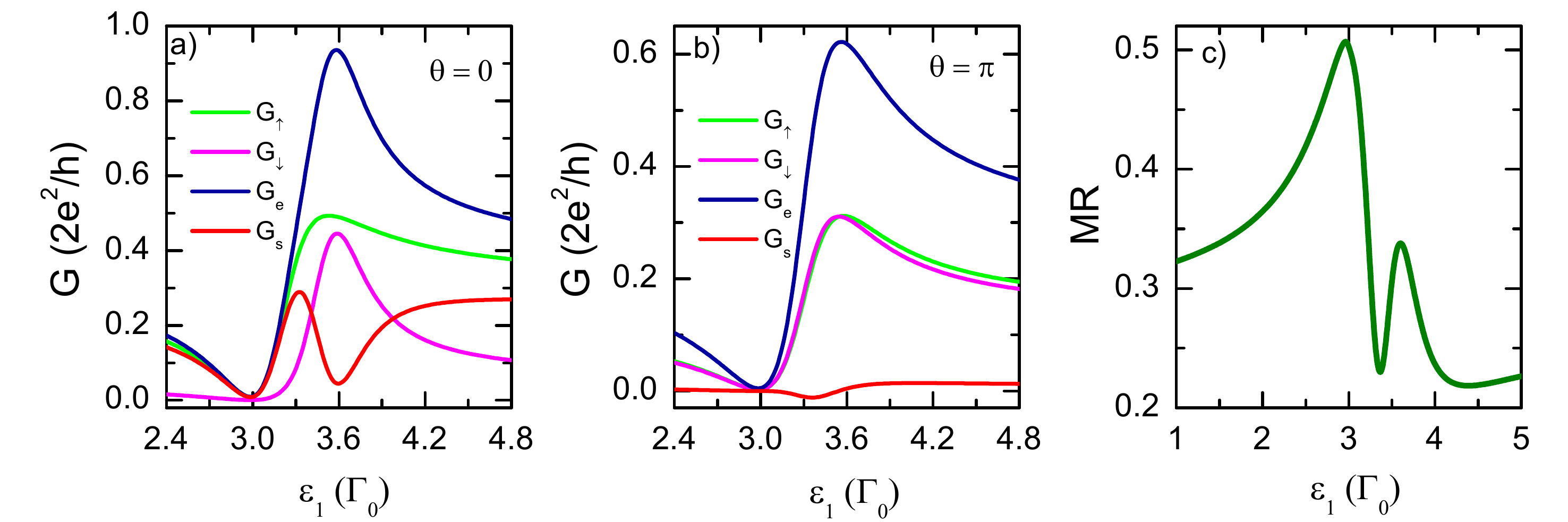}
\caption{a) G$_{\uparrow}$, G$_{\downarrow}$, along with G$_e$ and G$_s$ for the parallel configuration, as functions of the energy level of the auxiliary dot; b) the same quantities for the antiparallel configuration; c) the magneto-resistance, as defined in (\ref{mr}). Fixed parameters are $P = 0.6$, $k_BT = 0.03\ \Gamma_0$, $eV = 6 \ \Gamma_0$, $ \epsilon_2 =  3\ \Gamma_0 = eV/2 $, $U=0.9 \ \Gamma_0$ e $t_{12} = 0.65\ \Gamma_0$.}
\label{fig:magnetoresistance}
\end{figure*}

 So far, we have considered the dot levels as constants when studying the electric and spin conductance as functions of the applied voltage. Now, we discuss the effects of changing the relative positions of the dot levels by applying a gate voltage control to the auxiliary dot, thus varying $\epsilon_1$. 
 The other dot level $\epsilon_2$ is kept fixed and aligned with the chemical potential of the left electrode. We compare the parallel (Fig. \ref{fig:magnetoresistance}a) and antiparallel (Fig. \ref{fig:magnetoresistance}b) setups, noting that the electrodes are identical and with the same polarization (we only change the relative orientation of the magnetizations). All the cases plotted display an asymmetric Fano pattern, with a minimum and a maximum that result from path interference. The anti-resonance is pinned at the energy which is aligned with the chemical potential of the left lead, which acts as an infinite reservoir. In the parallel configuration, the resonance peaks are spin dependent due to the non-zero polarization. In the anti-parallel case, the spins are interchanged when going from one electrode to the other, thus avoiding the splitting of the spin dependent conductances. The small difference seen in Fig.~\ref{fig:magnetoresistance}b) between the up and down spins comes from the presence of a finite correlation $U$. The interaction also induces a negative spin conductance near the resonant peak. In order to assess the effect of changing the relative orientation of the ferromagnetic leads, one usually defines the magnetoresistance $MR$ as:
\begin{equation}\label{mr}
  MR=\frac{G_P-G_{AP}}{G_P}~,
\end{equation}
where G$_P$ and G$_{AP}$ are the conductances for the parallel ($\theta=0$) and anti-parallel ($\theta=\pi$) configurations, respectively. With definition (\ref{mr}), the maximum value of $MR$ is unity. In Fig.~\ref{fig:magnetoresistance}c), we show the magnetoresistance behavior corresponding to the cases displayed in panels a) and b), as a function of the energy level of the auxiliary dot. Peaks in $MR$ are observed at the position of the Fano resonances, with a maximum around the anti-resonance (more than $50\%$). By varying the gate voltage, one can tune the position of the $\epsilon_1$ level to one of those peaks, a feature that could be useful in applications.


\section{Conclusions} \label{Sec-conclusions}

Using Keldysh Green's function formalism, we have studied spin dependent transport properties in a T-shaped double quantum dot system with ferromagnetic electrodes. This setup is specially interesting, since spin currents are induced by ferromagnetism, and the T-shaped topology is responsible for the Fano effect. The interplay of both phenomena produces a variety of interesting responses, amenable of new spintronic applications. This is a clear advantage  of the present device in comparison with single-quantum dot systems.
Interference and quantum coherence yield resonances (peaks and dips) that can be used as filters for the current (electric and spin), by properly tuning gate voltages at the dots. We have also investigated the effects of the on-site Coulomb interaction $U$ in one of the dot (QD$_2$, central dot), which is assumed to be more correlated than the other (QD$_1$, side dot). A mean field approximation is used to handle the correlation and qualitatively understand the physics involved. Fano resonances dwindle with increasing values of $U$, when the correlation induces bound states at the side dot. The interaction is also associated with negative values of the spin conductance. Spin currents, which are only obtained for non-zero polarization, are dependent on the relative orientation of the lead magnetizations. With identical leads, in the anti-parallel configuration, the spin current vanishes in the absence of the Coulomb interaction. With $U\neq 0$, our mean field calculation suggests that a small spin current is generated with the voltage bias, in a limited interval of the voltage, where the spin conductance assumes negative values. This effect, if occurring in fact, may have the potentiality of interesting applications \cite{esaki}. We also found magnetoresistance effects when a gate voltage is applied to the side dot, with peaks around the position of resonances. \\

In summary, we have probed the transport properties of a hybrid nanodevice, with special focus on spin currents generation. The results shown in this paper can possibly be implemented in applications, since devices similar to our example can be fabricated using the current state of the art in the area.



\section*{Acknowledgements} \label{Sec-acknowledgements}

We acknowledge support from S\~ao Paulo Research Foundation (FAPESP) through project No. 2011/19298-4. GGC is grateful to Prof. Roberto Lagos for fruitful discussions.

\section*{References}

\bibliographystyle{elsarticle-num}
\bibliography{Referencias}

\end{document}